\documentclass[fleqn,twoside]{article}
\usepackage{amsmath,amssymb}
\usepackage{espcrc2}

\usepackage{graphicx}
\usepackage[figuresright]{rotating}

\title{ HMXB, ULX and star formation}

\author{M.Gilfanov\address[MPA]{Max-Planck-Institut f\"ur Astrophysik,
            85741 Garching b.M\"unchen, Germany}\address[IKI]{Space
            Research Institute, Profsoyuznaya 84/32, Moscow, Russia}, 
	H.-J.Grimm\addressmark[MPA]
	and
	R.Sunyaev\addressmark[MPA]\addressmark[IKI]}
       
\begin{document}

\begin{abstract}
Based on recent X-ray observations of the Milky Way,
Magellanic Clouds and nearby starburst galaxies we study
population of high mass X-ray binaries, their connection with 
ultra-luminous X-ray sources and relation to the star
formation. Although more subtle SFR dependent effects
are likely to exist, the data in the $\lg(L_X)\sim36-40.5$
luminosity range are broadly consistent with existence of a
``universal'' luminosity function of HMXBs, which can be roughly
described as a power law with differential slope of $\sim 1.6$ and a 
cutoff at $\lg(L_X)\sim 40.5$. 
The ULX sources found in many starburst galaxies occupy the high
luminosity end of this single slope power law distribution, whereas
its low luminosity part is composed of ``ordinary'' high mass X-ray
binaries, observed, e.g. in the Milky Way and Magellanic Clouds. 

As the normalization of the ``universal'' luminosity function is
proportional to the star formation rate, the number and/or the
collective X-ray luminosity of HMXBs can be used to measure the
current value of SFR in the host galaxy.
Distant (unresolved) starburst galaxies observed by Chandra at
redshifts of $z\sim 0.2-1.3$ obey the same $L_X-$SFR relation as local
galaxies, indicating that the ULXs at these redshifts were not
significantly more luminous than those found in nearby galaxies.   
\end{abstract}

\maketitle

\begin{figure*}[t]
\centerline{\hbox{
\resizebox{0.45\hsize}{!}{\includegraphics{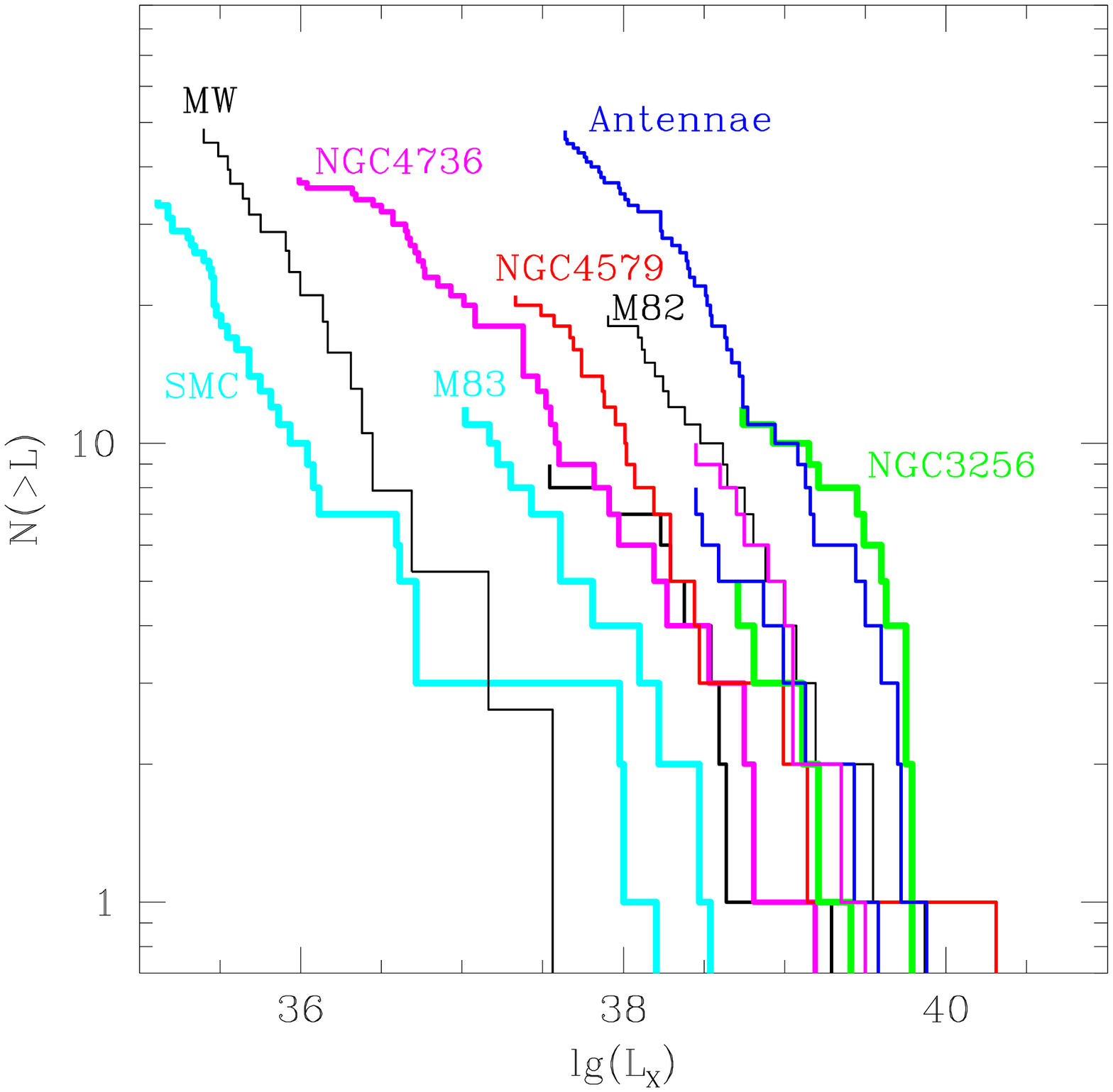}}
\resizebox{0.45\hsize}{!}{\includegraphics{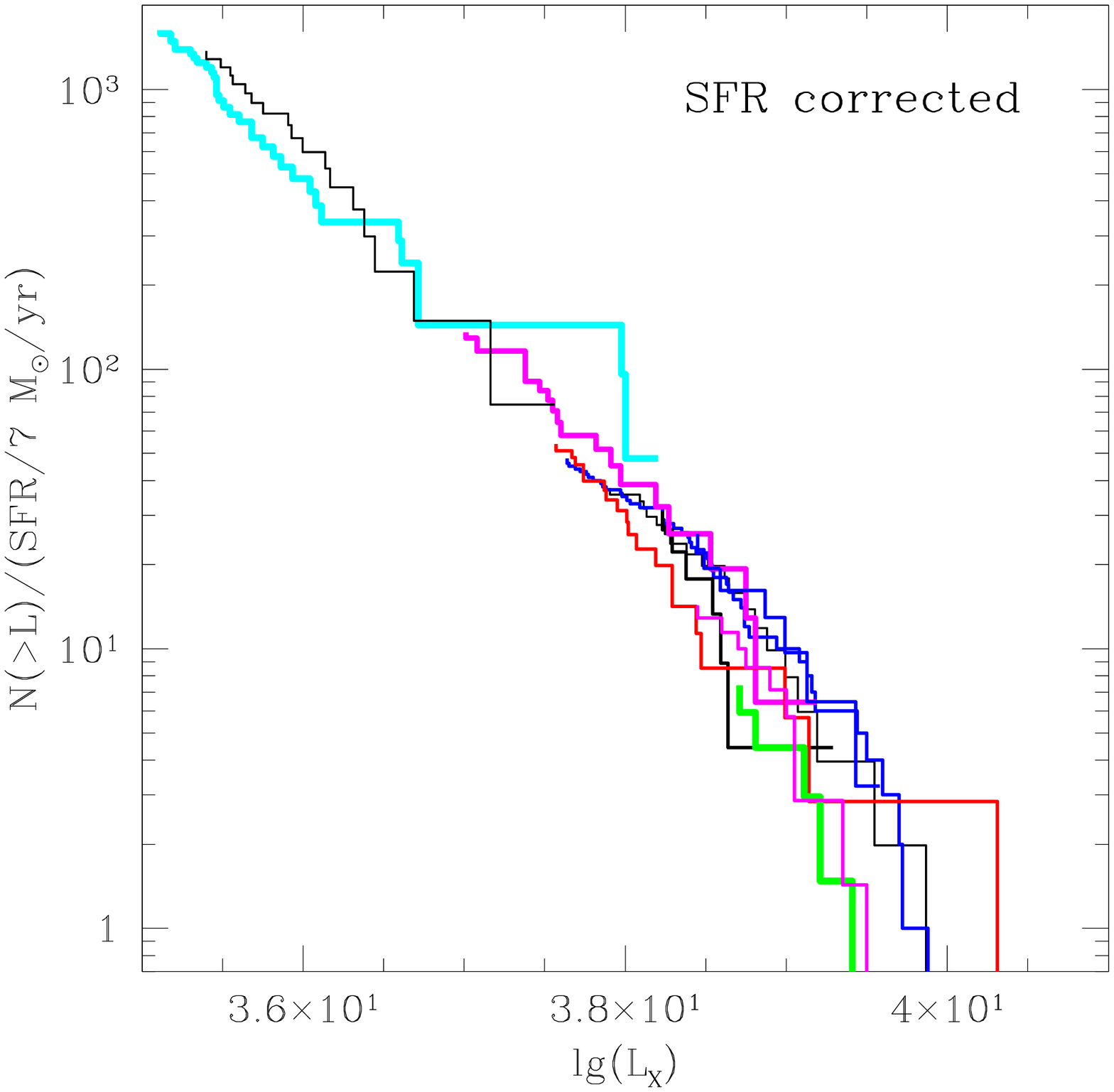}}
}}
\caption{{\em Left:} The luminosity functions of compact X-ray
sources in nearby galaxies obtained by Chandra. 
{\em Right:} The luminosity functions for the same galaxies scaled by
the ratio of their star formation rate to that of Antennae.}  
\label{fig:lfs}
\end{figure*}

\section{INTRODUCTION}

An unusual class of compact sources -- ultraluminous X-ray sources,
has been discovered in nearby galaxies more than a decade ago
\cite{colbert99,fab89}.  
Although bright, $L_X>10^{39}$ erg/s, point-like sources are found
both in young star forming galaxies and in old stellar population of
elliptical galaxies, the most luminous and exotic objects are
associated with actively starforming galaxies.  
Their nature and relation to more ordinary X-ray binaries  
is still a matter of a significant debate. Based on a  
simple Eddington luminosity argument, they appear to be powered by
accretion onto an intermediate mass object -- a black hole with the
mass in the hundreds-thousands solar masses range. However, a number
of alternative models have been considered as well -- from collimated 
radiation to  $\sim$stellar mass black holes, representing the high
mass tail of the standard stellar evolution sequence and accreting in
the near- or slightly super-Eddington regime.

Sub-arcsec angular resolution of Chandra observatory
opened a new era in studying population of compact sources in nearby
galaxies. For the first time an opportunity was presented to observe
compact sources in a nearly confusion free regime and to investigate
their relation to fundamental parameters of the host galaxy, 
such as its stellar mass and star formation rate. 
Depending on the mass of the optical companion, X-ray binaries are
subdivided in to two classes -- low  and high mass X-ray 
binaries, having significantly different evolutionary time
scale, $\sim 10^{6-7}$ and $\sim 10^{9-10}$ years respectively 
\cite{xrbrev}. 
The prompt emission from HMXBs  makes them a potentially  good tracer
of the very recent star formation activity in the host galaxy
\cite{rs78}.  The LMXBs, on the other hand, have no relation to the 
present star formation, but, rather, are related to the stellar
content of the host galaxy \cite{lmxb}.
Chandra observations of the nearby galaxies present a possibility to
verify this simple picture and to calibrate the relation between the
HMXB population and the current value of SFR. This
opens a new way to determine the star formation rate in relatively
closeby galaxies as well as in young and very distant ones. 
Detailed study of the population of the compact sources in star
forming galaxies can also shed the light on the mistery of the ULXs.

\begin{figure*}[t]
\centerline{\hbox{
\resizebox{0.45\hsize}{!}{\includegraphics{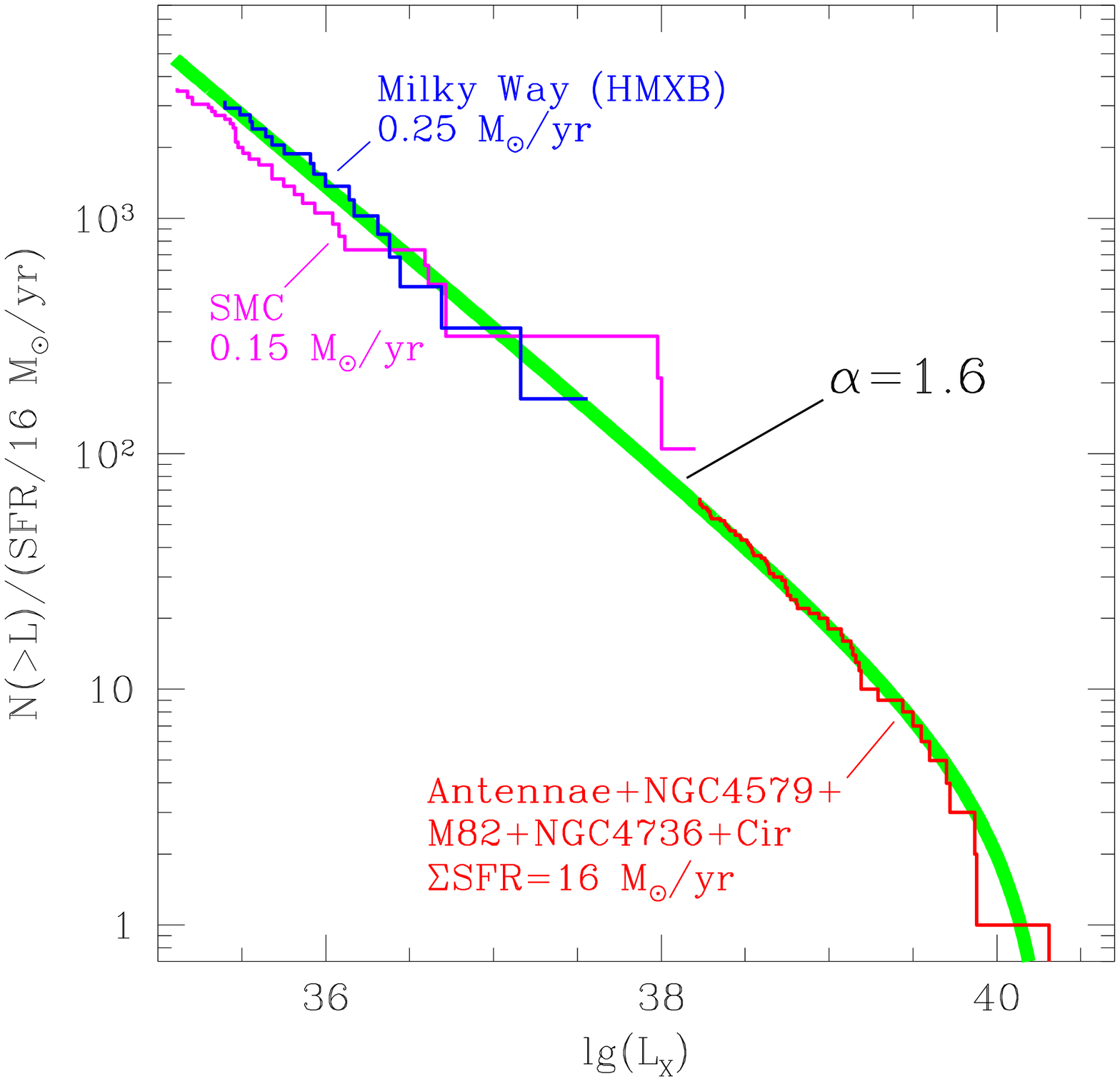}}
\resizebox{0.45\hsize}{!}{\includegraphics{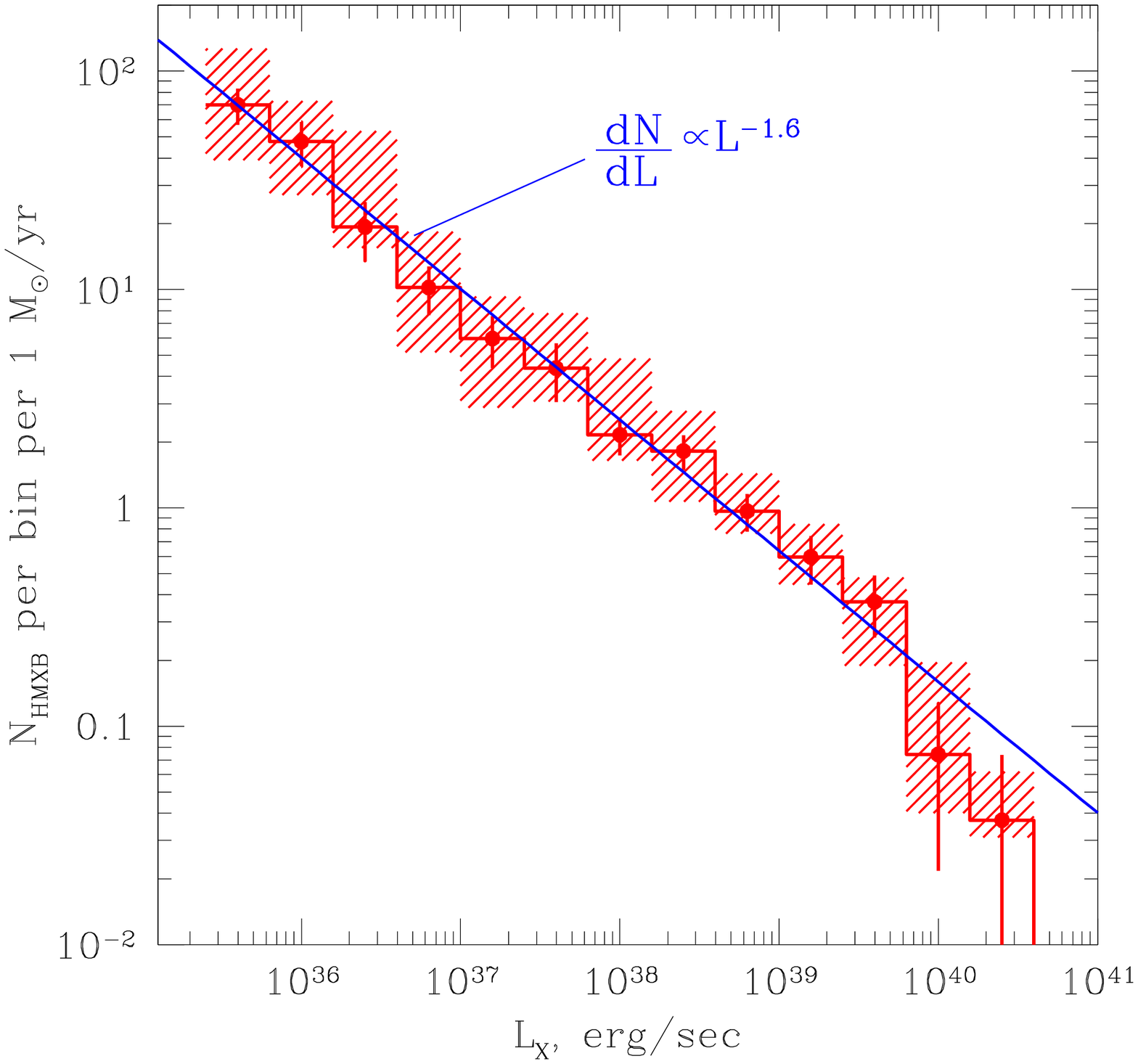}}
}}
\caption 
{{\em Left:} Combined luminosity function of compact X-ray   sources
in starburst galaxies and the luminosity functions of HMXBs in the
Milky Way and SMC. The thick grey line is the best
fit to the starburst galaxies, Eq.(\ref{eq:xlfdif}). 
{\em Right:} Average differential luminosity function of the galaxies
from our Chandra sample. The straight line is a power law 
defined by eq.(\ref{eq:xlfdif}).  
The shaded area illustrates the amplitude of systematic errors (90\%
confidence level) associated with  uncertainties in the adopted SFR
values (assuming 30\% relative error) and the distances (20\% relative
error).
}   
\label{fig:lf_comb}
\end{figure*}

\section{``UNIVERSAL'' LUMINOSITY FUNCTION OF HMXBs}

In order to study the HMXB population and its relation to
star formation we selected a number of nearby late type galaxies
observed by Chandra, based primarily on two criteria. 
(i) The galaxy can be spatially resolved 
by Chandra, so that the contribution of a central AGN
can be discriminated and the luminosity functions of the compact
sources can  be constructed without severe confusion effects.
(ii) The galaxy has sufficiently high 
SFR/mass ratio, to ensure that the population of X-ray binaries is
dominated by HMXBs and the LMXB contribution of can be safely ignored
\cite{lx-sfr}.  For the Milky Way we explicitly selected HMXBs, based
on results of \cite{grimm1}.
For each galaxy, the star formation rate was determined
combining the results from conventional SFR indicators, \cite{sfrrev}
(FIR, UV, H$_{\alpha}$ and radio). The SFR values in the sample
range from $\sim 0.15$ to $\sim 7$ M$_{\odot}$/yr.
Further details and references to the X-ray data are given in
\cite{grimm2}.    

The Fig.\ref{fig:lfs} (left panel) shows observed luminosity
functions. They are characterized by a large spread in the
number of sources and in the luminosity of the brightest
source. However, after being rescaled to the
same value of the star formation rate, they appear to match each other 
both in the slope of 
the distribution and the normalization (right panel in
Fig.\ref{fig:lfs}). Although a finite dispersion might still exist,
especially at the high luminosity end, the rescaled luminsity
functions occupy a rather narrow band in the $N-L_X$ plane, despite of
large dynamical range of the star formation rates, a factor of 
$\sim 50$. 
We further combine the data for five nearby starburst galaxies with
the best known luminosity functions,  having cumulative star fromation 
rate of $\approx 16$ M$_{\odot}$/yr, and compare them with the
luminosity distributions of two low SFR galaxies, the Milky Way and
SMC (Fig.\ref{fig:lf_comb}). As previously, a good agreement both in
the slope and normalization of luminosity distributions is obvious. 
The fit to combined data with a power law distribution gives the slope
of $\alpha\approx 1.6$, a cut-off at $L_{\rm cut}\approx (2-3)\cdot
10^{40}$ erg/s and the normalization proportional to the star
formation rate: 
\begin{eqnarray}
\frac{dN}{dL}=(3.3^{+1.1}_{-0.8})\cdot SFR \cdot L_{38}^{-1.61\pm0.12}
\label{eq:xlfdif}
\end{eqnarray}
where SFR is formation rate for massive stars, $M>5M_{\odot}$. Within
the accuracy of the present analysis, this can be regarded as a
``universal'' luminosity function of high mass X-ray binaries in
nearby star forming galaxies. The Fig.\ref{fig:lf_comb} shows the
average differential luminosity function of HMXBs obtained using the
data of all galaxies from our Chandra sample.

\begin{figure*}[t]
\centerline{\hbox{
\resizebox{0.45\hsize}{!}{\includegraphics{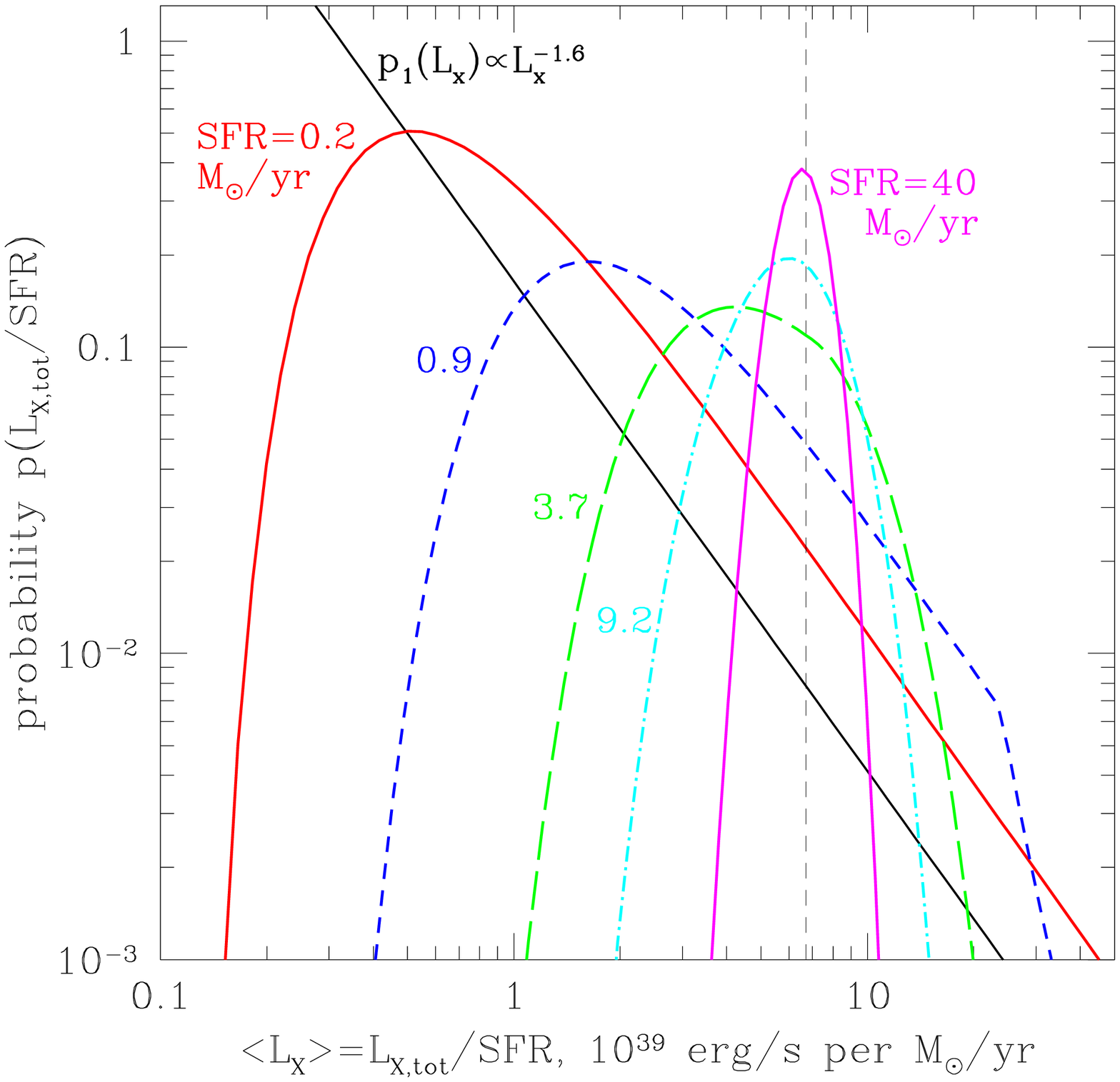}}
\resizebox{0.45\hsize}{!}{\includegraphics{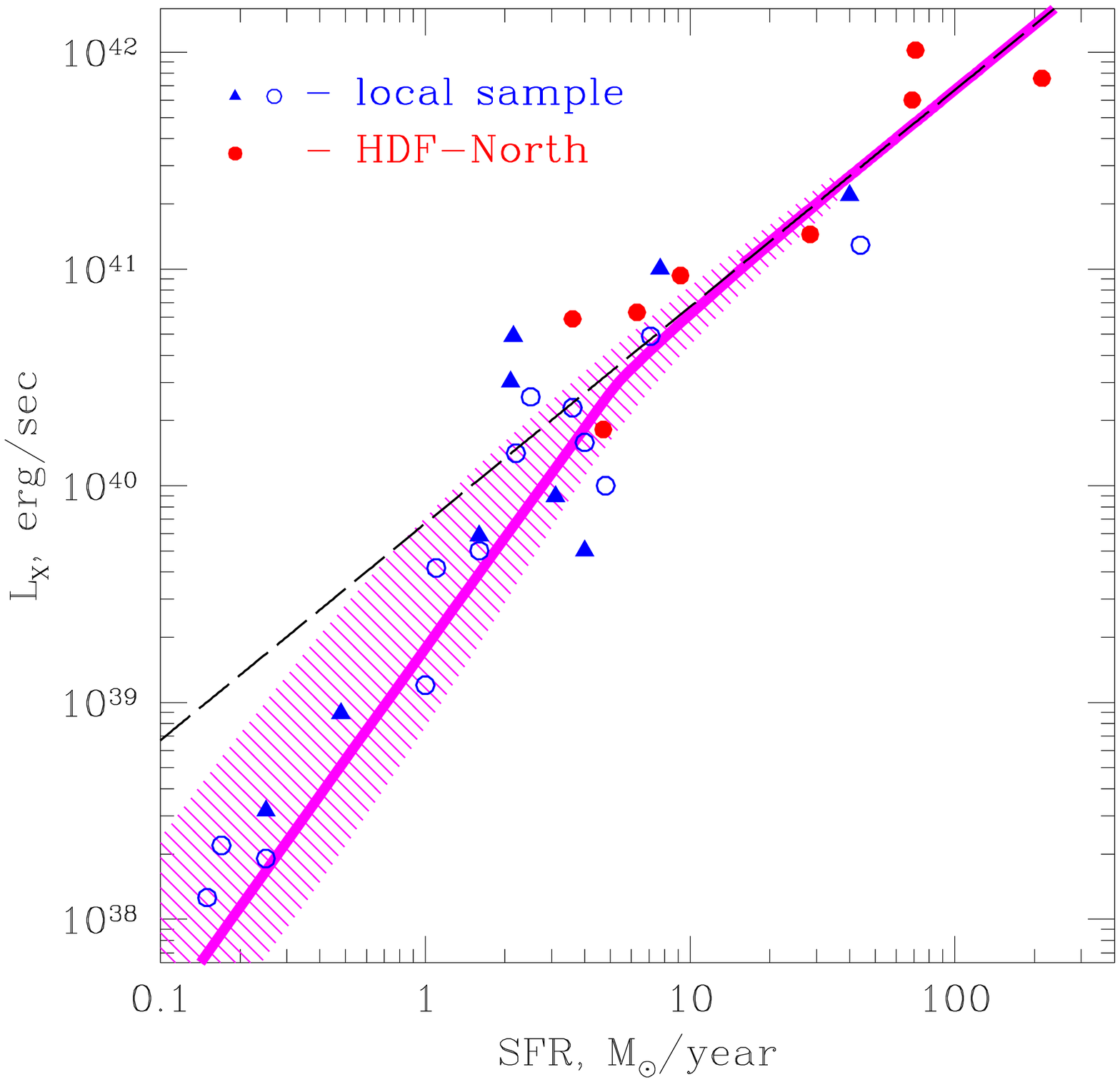}}
}}
\caption{
{\em Left:} The probability distributions $p(L_{\rm tot}/{\rm SFR})$
for different values of SFR. The vertical dashed line shows the
expectation mean, defined by eq.(\ref{eq:ltot_mean}).
{\em Right:} The $L_X-$SFR relation. The open circles are nearby 
galaxies observed by Chandra, the filled triangles are nearby galaxy
observed by ASCA and BeppoSAX, for which only total luminosity is
available, the filled circles are distant star
forming galaxies from  the Hubble Deep Field North. The thick grey
line is relation between the star formation rate and the most probable
value of the total luminosity, predicted from the ``universal''
luminosity function of HMXBs, the shaded area shows 67\% intrinsic
spread of the $L_X-$SFR relation, the dashed line is the expectation
mean, defined by eq.(\ref{eq:ltot_mean}).} 
\label{fig:lx-sfr}
\end{figure*}

\section{HIGH MASS X-RAY BINARIES AS A STAR FORMATION INDICATOR}

As the population of high mass X-ray binaries (i.e. the normalization
of the luminosity function and the number of sources) is proportional
to SFR, eq.(\ref{eq:xlfdif}), they can be exploited to measure star 
formation rate in the host galaxy. Such a method, based on the X-ray
emission of a galaxy, might circumvent one the main sources of
uncertainty of conventional SFR indicators -- absorption by dust and
gas \cite{sfrrev}. Indeed, galaxies are mostly transparent to X-rays
above $\sim 2$ keV, except for the densest parts of the most massive
molecular clouds. 

For nearby, spatially resolved galaxies, the star formation rate can
be determined via a direct fitting of the luminosity distribution of
the compact sources or counting the number of sources above a given
luminosity threshold. For distant, unresolved galaxies one can use
the total luminosity of the galaxy.  

There are, however, two main complicating factors in using the X-ray
luminosity of a galaxy as a SFR indicator: (i) X-ray emission from the
central supermassive black hole, which even in the case of a low
luminosity AGN can easily out-shine X-ray binaries. This does not 
present a problem in nearby ($D<20-30$ Mpc) galaxies as the Chandra
resolution is sufficient to separate contribution of the central
source, but might become an important source of contamination in
spatially unresolved galaxies. 
(ii) Contribution of low mass X-ray binaries. This equally affects
nearby and distant galaxies, as LMXBs can not be easily separated from
HMXBs using the X-ray data in the Chandra bandpass, and the optical
identifications are (potentially) available for the most nearby
galaxies only. However, as the LMXB population scales linearly with
the stellar mass of the host  galaxy \cite{lmxb}, their contribution
can be estimated and, potentially, corrected for. Obviously, at
sufficiently high values of SFR/$M_*$ it can be safely neglected,
\cite{lx-sfr}.

To further calibrate the $L_X-$SFR relation and verify its
applicability to distant galaxies, we extended our sample to include
the spatially unresolved galaxies observed by ASCA and BeppoSAX and
distant star forming galaxies detected by Chandra in the Hubble Deep
Field North \cite{hdfn} (see \cite{grimm2} for details and
references). As for the 
galaxies from our spatially resolved  Chandra sample, star formation
rates  were determined by the conventional, non X-ray methods. The
combined data are shown in the $L_X-$SFR plane in
Fig.\ref{fig:lx-sfr}.

\subsection{Effects of statistics and \boldmath $L_X-$SFR relation}

A seemingly obvious expression for the total luminosity can be
obtained integrating the luminosity distribution (\ref{eq:xlfdif}):
\begin{eqnarray}
\left  <L_{\rm tot} \right > =
\int_{L_{\rm min}}^{L_{\rm cut}} \frac{dN}{dL}\,L\,dL
\propto SFR
\label{eq:ltot_mean}
\end{eqnarray}
implying, that the total luminosity is proportional to the
star formation rate. We note, however, that the quantity of interest
is a sum of the luminosities of discrete sources:
\begin{eqnarray}
L_{\rm tot}=\sum_{k} L_k
\end{eqnarray}
with $L_k$ obeying a power law probability distribution given by
eq.(\ref{eq:xlfdif}). For a sufficiently flat slope, $\alpha<2$, the 
total luminosity $L_{\rm tot}$ will be defined by the brightest
sources, corresponding to the high luminosity end of the power law
distribution eq.(\ref{eq:xlfdif}). 
Their actual number in a galaxy will obey the Poisson distribution
$P(n,\mu)=\mu^ne^{-\mu}/n!$ (with $\mu\propto$SFR), 
which for small values of $\mu$ is significantly asymmetric. 
As a consequence, for a galaxy with  small SFR, the probability
distribution $p(L_{\rm tot})$ will be also strongly asymmetric, as
illustrated by the left panel in Fig.\ref{fig:lx-sfr}.  
Because of its skewness, the mode of the $p(L_{\rm tot})$ distribution
-- the value of $L_{\rm tot}$ that would be most likely measured in an
arbitrarily  chosen galaxy, is not equal to the expectation mean
defined by eq.(\ref{eq:ltot_mean}).  
Only in the large SFR limit, when there are sufficiently many sources
with luminosities $L\sim L_{\rm cut}$, these 
two quantities become close to each other.\footnote{Obviously in the
case of e.g. flat ($dN/dL=$const) or Gaussian  flux distribution the
most probable value of $L_{\rm tot}$ always equals to the expectation
mean defined by  eq.(\ref{eq:ltot_mean}). 
}

The difference between these two quantities is further illustrated in
the right panel in Fig.\ref{fig:lx-sfr}, showing the predicted
$L_X$--SFR relation, calculated using the parameters of the
``universal'' HMXB luminosity function, eq.(\ref{eq:xlfdif}). 
The solid line in  the figure  shows the SFR--dependence of the 
mode of the probability distribution $p(L_{\rm tot})$ and predicts the
{\em most probable} value of the X-ray  luminosity of a randomly
chosen galaxy.  
If observations of many (different) galaxies with close values of SFR
are performed, the obtained values of $L_{\rm tot}$ will obey the
probability distribution depicted in the left panel of 
Fig.\ref{fig:lx-sfr}. The average of the measured values of 
$L_{\rm tot}$ will be equal to the expectation mean given by
eq.(\ref{eq:ltot_mean}) and shown by the dashed straight lines in the
left and right panels of Fig.\ref{fig:lx-sfr}. 
Due to the properties of the probability  distribution 
$p(L_{\rm tot})$ these two quantities are not identical in the low SFR
limit, when the total luminosity is defined by a small number of the
most luminous sources.

\subsection{\boldmath $L_X-$SFR relation: predicted vs. observed}

The right panel in Fig.\ref{fig:lx-sfr} compares the data with the
predicted $L_X-$SFR relation. Good agreement, both in the non-linear
low SFR regime and at high SFR values is apparent. The relations
between total X-ray luminosity of a galaxy due to HMXBs and the star
formation rate are:
\begin{equation}
{\rm SFR} [{\rm M}_{\odot}/{\rm yr}] \approx \frac{L_{\rm
2-10~keV}}{6.7\cdot 10^{39}{\rm ~erg/s}}
\label{eq:sfr_lx}
\end{equation}
in the linear regime and
\begin{equation}
{\rm SFR} [{\rm M}_{\odot} {\rm yr}^{-1}] \approx \left(\frac{L_{\rm
2-10~keV}}{2.6\cdot 10^{39}{\rm~erg/s}}\right)^{0.6}
\label{eq:sfr_lx1}
\end{equation}
in the non-linear regime, corresponding to SFR$< 4.1$
M$_{\odot}$ yr$^{-1}$ ($L_{\rm 2-10~keV}< 2.8\cdot 10^{40}$ erg s$^{-1}$)
The former, linear, regime of the $L_X$--SFR
relation was studied independently by \cite{ranalli} based on ASCA
and BeppoSAX data. Note that their equation (12) agrees with our
eq.(\ref{eq:sfr_lx}) within $\sim 30$ per cent.

Due to skewness of the probability distribution $p (L_{\rm tot})$,
large and asymmetric dispersion around the solid curve in
Fig.\ref{fig:lx-sfr} is expected in the non-linear low SFR regime.  
This asymmetry is already seen from the distribution of the points  in
Fig.\ref{fig:lx-sfr} -- at low SFR values there are more points
above the solid curve, than below. Moreover, the galaxies
lying significantly above the solid and dashed curves in
Fig.\ref{fig:lx-sfr} should be expected at low SFR and will
inevitably appear as the plot is  populated with more objects.  
Such behavior differs from a typical
astrophysical situation and  should
not be ignored when analyzing and fitting the $L_X$--SFR relation in
the low SFR regime. In particular, the standard data analysis
techniques -- least square and $\chi^2$ fitting become inadequate.

\begin{figure}[t]
\resizebox{\hsize}{!}{\includegraphics{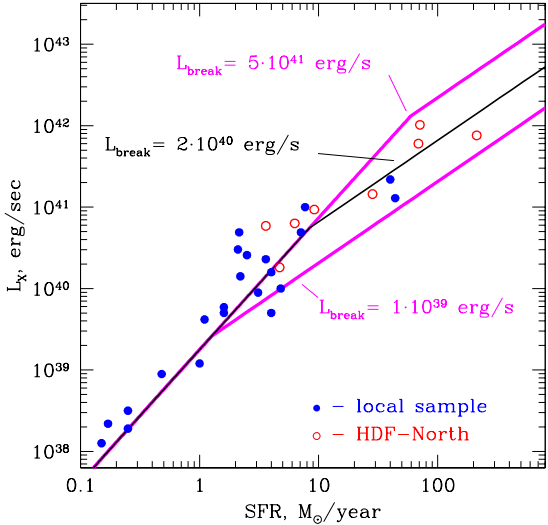}}
\caption{Dependence of the $L_X-$SFR relation on the maximum
luminosity of the sources, $L_{\rm cut}$. The three curves,
corresponding to different values of $L_{\rm cut}$ coincide in the
non-linear low SFR regime but differ in the position of the break
between linear and non-linear regimes.
The data are the same as in Fig.\ref{fig:lx-sfr}.
}
\label{fig:lx-sfr-break}
\end{figure}

\section{DISCUSSION}

\subsection{High luminosity cut-off in the HMXB luminosity function}

The existence of the linear regime in the $L_X$--SFR relation is a
direct consequence of the cut-off in the luminosity function. 
The position of the break between non-linear and linear parts of the
$L_X$--SFR relation depends on the slope of the luminosity function
and the value of the cut-off luminosity (Fig.\ref{fig:lx-sfr-break}):
${\rm SFR}_{\rm break}\propto L_{\rm cut}^{\alpha-1}$. This allows one
to constrain parameters of the luminosity distribution of compact
sources using the data of spatially unresolved galaxies.

Agreement of the predicted $L_X-$SFR relation with the data both
in high and low SFR regimes gives an independent confirmation of the
existence of a cut-off in the luminosity function of HMXBs at $L_{\rm
cut}\sim {\rm several} \times 10^{40}$ erg/s
(Fig.\ref{fig:lx-sfr},\ref{fig:lx-sfr-break}).  
It also confirms that $L_X-$SFR data, including the high redshift
galaxies from Hubble Deep Field North, are consistent with the HMXB
luminosity function parameters, derived from significantly fewer
galaxies, than plotted in 
Figs.\ref{fig:lx-sfr},\ref{fig:lx-sfr-break}.

\subsection{Ultraluminous X-ray sources}

One of the surprising results of this study is a smooth, single
slope power law shape of the average luminosity function of 
compact sources in star forming galaxies, without any
significant steps and features in a broad luminosity range,
$\lg(L_X)\sim 36-40.5$ (Fig.\ref{fig:lf_comb}).
The high luminosity end, $\lg(L_X)>39$, of this distribution 
corresponds to ultraluminous X-ray sources. 
Its low luminosity end, on the other hand, is composed of ordinary
X-ray binaries, powered by accretion onto a $\sim$stellar mass compact
objects.  
This result  constrains the range of  possible models for ULXs. 
Their frequency and luminosity distributions should be a smooth
extension towards higher luminosities of that of ``ordinary''
$\sim$stellar mass systems, emerging from the standard stellar
evolution sequence.   
Although some of the ULXs might be indeed rare and exotic objects,
it appears that majority of them cannot be a completely
different type of the source population, but, rather,  represent the
high mass, high $\dot{M}$ tail of the HMXB population.

The luminosity of ULXs in the nearby galaxies has a maximum value of
the order of $\lg(L_X)\sim 40.5$. The fact that the galaxies from the
Hubble Deep Field North obey the same $L_X-$SFR relation
(Fig.\ref{fig:lx-sfr-break}), implies, that the ULXs at the redshift
of $z\sim 0.2-1.3$ were not significantly more luminous, that those
observed in the nearby galaxies.

\begin{figure*}[t]
\centerline{\hbox{
\resizebox{0.45\hsize}{!}{\includegraphics{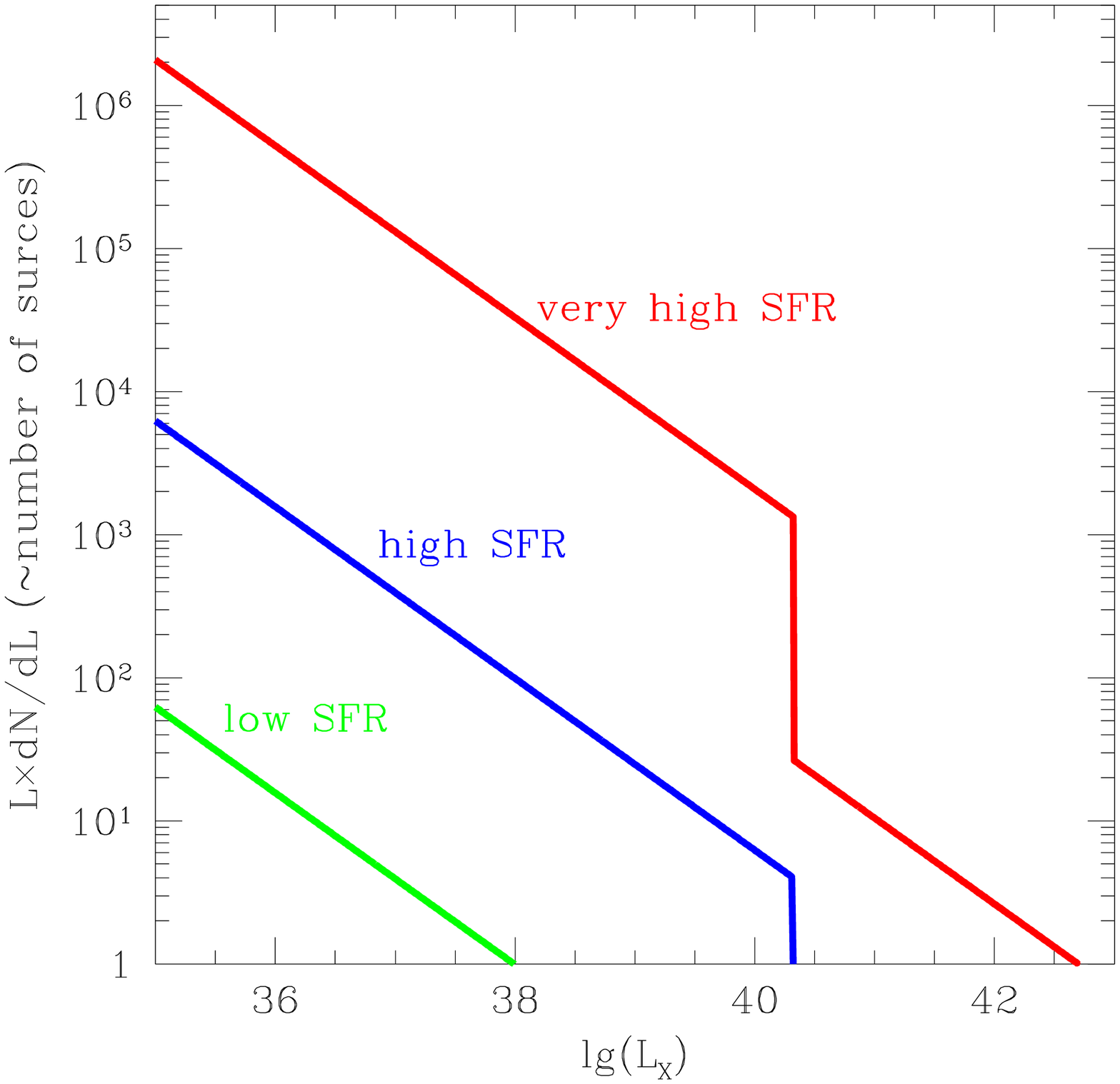}}
\resizebox{0.45\hsize}{!}{\includegraphics{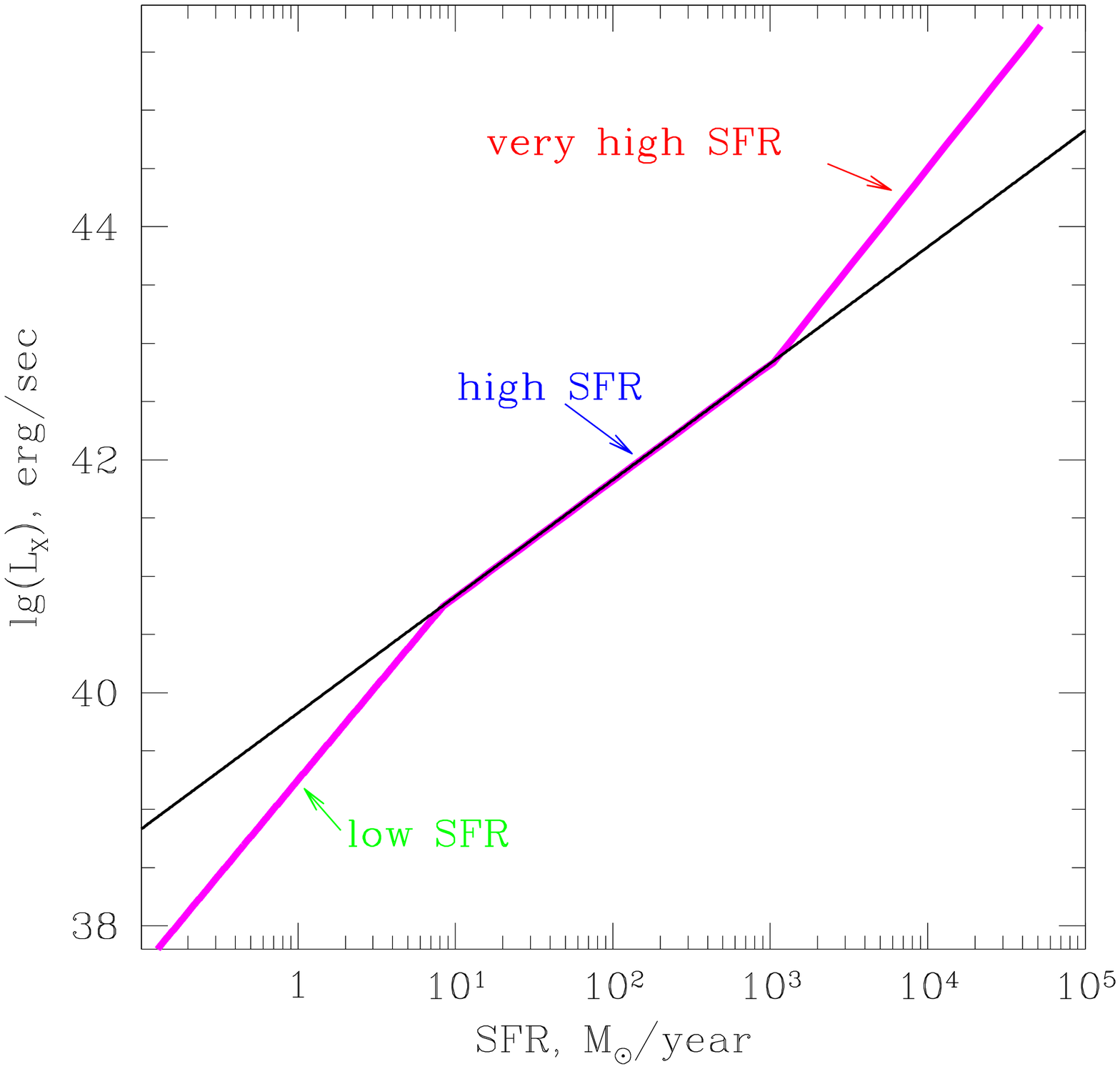}}
}}
\caption{Illustration of the effect of hypothetical intermediate mass
black holes on the $L_X-$SFR relation. 
{\em Left:} The luminosity function of compact sources at different
levels of star formation rate. 
{\em Right:} Corresponding $L_X-$SFR relation. The thin straight line
shows the linear relation.
\label{fig:imbh}}
\end{figure*}

\subsection{Intermediate mass black holes}

The hypothetical intermediate mass black holes, probably reaching
masses of $\sim 10^{2-5} M_{\odot}$, might be produced, e.g. via black 
hole merges in the dense stellar clusters, and can be associated with 
extremely high star formation rates. To accrete efficiently,
they should form close binary systems with normal
stars or be located  in dense molecular clouds. It is natural to
expect, that such objects are  significantly less frequent than
$\sim$stellar mass black holes. 
The transition from the $\sim$stellar mass BH HMXB to intermediate
mass BHs should manifest itself as a step in the luminosity
distribution of compact sources (Fig.\ref{fig:imbh}, left panel).  
If the cut-off in the HMXB luminosity function, observed at 
$\lg(L_{\rm cut}) \sim 40.5 $ corresponds to the maximum
possible luminosity of $\sim$stellar mass black holes and if at
$L>L_{\rm cut}$ a population of 
hypothetical intermediate mass BHs emerges, it should lead to
a drastic change in the slope of the $L_X$--SFR relation at extreme
values of SFR \cite{stat} (Fig.\ref{fig:imbh}, right panel). 
Therefore, observations of distant star forming galaxies with very
high SFR might be an easy way to probe the population of
intermediate mass black holes.


\end{document}